\documentclass[aps,twocolumn]{revtex4}
\usepackage{graphicx}
\usepackage{amssymb}
\usepackage{amsmath}
\usepackage{bm}
\usepackage{color}

\def\ra{\rangle}
\def\la{\langle}

\def\e{\varepsilon}

\def\s{\sigma}
\def\t{\theta}
\def\bk{{\bm k}}
\def\M{{\bm {\hat M}}}

\begin{document}

\title{Phase diagram of a magnetic topological nodal semimetal:
	\\ Stable nodal line in an easy-plane ferromagnet}
\author{Yuya Ominato$^{1}$, Ai Yamakage$^{2}$, and Kentaro Nomura$^{1,3}$}
\affiliation{Institute for Materials Research, Tohoku university, Sendai 980-8577, Japan}
\affiliation{Department of Physics, Nagoya University, Nagoya 464-8602, Japan}
\affiliation{Center for Spintronics Research Network, Tohoku University, Sendai 980-8577, Japan}
\date{\today}

\begin{abstract}
We  study a topological phase diagram of a ferromagnetic topological nodal semimetal.
We consider a lattice model for three dimensional topological insulators with ferromagnetic ordering.
The exchange coupling between the magnetization and the electron spin leads to the nodal band structure.
The topology of the nodal band structure depends on the direction of the magnetization and both the Weyl points and the nodal line emerge.
We find that the nodal line structure is stable under an easy-plane magnetization in appropriate model parameters.
In this case, the nodal line phase emerges as a phase boundary between two topologically distinct Weyl semimetal phases.
\end{abstract}

\maketitle

\section{Introduction}
\label{sec_intro}

Topological semimetals, such as Weyl semimetals \cite{wan2011topological,burkov2011weyl,armitage2018weyl} and nodal line semimetals \cite{kim2015dirac,fang2015topological,yamakage2016line,chan2016ca}, have the nodal band structure and they are characterized by topological invariants.
Weyl semimetals have pairs of nodal points called Weyl points and Chern number characterizes the nodal structure. 
Nodal line semimetals have gap closing lines, where the conduction and valence bands stick together on one-dimensional lines in momentum space. Nodal lines are characterized by the Zak phase \cite{zak1989berry}.
Weyl points are stable against perturbations, while stable nodal lines require additional symmetry.
If spin-orbit coupling is negligible, the nodal line is protected by the composition of inversion and time-reversal symmetry \cite{burkov2011topological,kim2015dirac,fang2015topological} or crystalline symmetries such as the mirror-reflection symmetry \cite{yamakage2016line,chan2016ca} and the glide-plane symmetry \cite{huang2017topological,shao2018nonsymmorphic}. Under these symmetry conditions, the Zak phase is $Z_2$ quantized.
However, spin-orbit coupling gaps out the nodal line and leads to either topological insulator or Dirac semimetal phases \cite{kobayashi17}. There is only a few candidate materials, where the nodal lines are stable even in the presence of spin-orbit coupling \cite{kim2015surface,bian2016topological, liang16, du2017emergence, sun17, nie2019topological, funada19}.

At the present moment, there are many experimental observations of Weyl semimetals \cite{xu2015discovery,lv2015experimental,nakatsuji2015large,nayak2016large,liu2018giant,wang2018large} and nodal line semimetals \cite{hu2016evidence,takane2016dirac,takane2018observation,hosen2018discovery,takane18} including non-magnetic and magnetic materials. Most of them are non-magnetic and a few of them are magnetic. There have been many theoretical proposals for magnetic topological semimetals \cite{wan2011topological,burkov2011weyl,xu2011chern,kurebayashi2014weyl,ueda2015magnetic,yang2017topological,jin2017ferromagnetic,xu2018topological,ominato18,nie2019topological}. 
In these systems, the nodal structures depend on the magnetic configuration, i.e. the topological properties of them are manipulated by the magnetization.
As we mentioned above, the stable nodal lines require additional symmetry.
In the case of $\beta$-${\rm V}_2{\rm OPO}_4$ \cite{jin2017ferromagnetic} and ferromagnetic rare-earth-monohalides \cite{nie2019topological}, the nodal lines are protected by mirror-reflection symmetry perpendicular to the magnetization direction. When the magnetization is tilted, the nodal lines are gapped and the Weyl semimetal phase emerges.

In this work, we study the nodal band structure of a ferromagnetic nodal semimetal. The nodal structure is modulated by changing the magnetization direction. In our model, there are orbital-independent and dependent exchange coupling terms, which we call $J_0$ and $J_3$ terms, respectively. 
The topological phase diagram is significantly different in the situation that the $J_0$ term is dominant and the $J_3$ term is dominant. When the $J_0$ term is dominant, the topological phase diagram consists of two topologically distinct Weyl semimetal phases and the nodal line semimetal phase. The two Weyl semimetal phases are distinguished by a sign of the Chern number. Changing the magnetization direction from the south hemisphere to the north, the sign of the Chern number changes at the equator. When the magnetization is in the $x$-$y$ plane, the system becomes the nodal line semimetal and the nodal line is stable. When the $J_3$ term is dominant, the nodal line semimetal phase emerges in the situation that the magnetization is pointing along $z$ axis. Once the magnetization is tilted from $z$ axis, the system becomes the Weyl semimetal phase.

The paper is organized as follows.
In Secs. \ref{sec_model} and \ref{invariants}, we introduce a model Hamiltonian and topological invariants. 
In Sec.\ \ref{sec_numerical}, we numerically calculate the Chern number and the Zak phase. 
Symmetry and the stable nodal line for in-plane magnetization are discussed in Sec. \ref{sec_discussion}.
Finally,
we summarize our result in Sec. \ref{sec_conclusion}.


\section{Model Hamiltonian}
\label{sec_model}

We consider a magnetically doped topological insulator. The model Hamiltonian is given as
\begin{align}
H_\bk=H_0+H_{\rm ex},
\end{align}
where $H_0$ is a lattice model for three dimensional topological insulators on a cubic lattice \cite{zhang2009topological,liu2010model},
\begin{align}
H_0=\tau_x\s_x t\sin{k_ya}-\tau_x \s_y t\sin{k_xa}+\tau_y t\sin{k_za}+m_\bk \tau_z,
\end{align}
where $a$ is a lattice constant, $t$ is a hopping parameter, and $m_\bk$ is a mass term,
\begin{align}
m_\bk&=m_0+m_2\sum_{\alpha=x,y,z}(1-\cos k_\alpha a).
\end{align}
$\bm{\s}$ and $\bm{\tau}$ are the Pauli matrices acting on the real spin and the pseudo spin (orbital) degrees of freedom.
The exchange coupling between the magnetization and the electron spin is written as \cite{wakatsuki2015domain}
\begin{align}
    H_{\rm ex}=J_0{\M}\cdot{\bm \s}+J_3\tau_z{\M}\cdot{\bm \s},
\end{align}
where $J_0$ and $J_3$ are exchange coupling constants and $\M$ is a unit vector representing the direction of the magnetization,
\begin{align}
\M=(\sin\t\cos\phi,\sin\t\sin\phi,\cos\t),
\end{align}
as depicted in Fig.\ \ref{fig_chern} (a).
There are two kinds of exchange terms, which we call the $J_0$ and $J_3$ terms, respectively.
The existence of the $J_3$ term originates from inequality of the exchange coupling constants between two orbitals considered here, i.e., $p$-orbitals of (Bi,Sb) and Te. In magnetic topological insulators, ${\rm Cr}_x({\rm Bi}_{1-y}{\rm Sb}_y)_{2-x}{\rm Te}_3$ \cite{chang2013experimental,checkelsky2014trajectory,kou2014scale}, Cr atoms are substituted for Bi or Sb atoms.
This leads to the inequality of exchange coupling between $p$-orbitals of (Bi,Sb) and Te.
We assume that $J_0\geq0$ and $J_3\geq0$. This assumption does not change the essential results in the following sections.

\section{Topological Invariants}
\label{invariants}

In the present model, both the Weyl points and the nodal line emerge.
The nodal structures are characterized by topological invariants.
In the following, subscripts $\alpha,\beta,\gamma$ refer to $x,y,z$.
The Weyl points are characterized by the Chern number defined as a function of $k_\alpha$ as
\begin{align}
C(k_\alpha)=\sum_{n}^{occ.}\iint\frac{dk_\beta dk_\gamma}{2\pi}\left[\nabla_\bk\times\bm{A}_n(\bk)\right]_\alpha,
\label{eq_chern}
\end{align}
where $k_\beta$ and $k_\gamma$ are wave numbers perpendicular to $k_\alpha$ and the summation is over the occupied states.
The Berry connection is defined as
\begin{align}
\bm{A}_n(\bk)=-i\la u_{n\bk}|\nabla_\bk|u_{n\bk}\ra,
\end{align}
where $|u_{n\bk}\ra$ is an eigenstate of the Hamiltonian $H_\bk$.
At a fixed $k_\alpha$, the system is regarded as a two-dimensional insulator with an integer Chern number. When one sweeps $k_\alpha$, the Chern number changes at the Weyl points because they behave as a monopole with positive or negative charge. One can assign monopole charge for each Weyl point by the change of the Chern number.
The nodal line is characterized by the Zak phase defined as a function of $k_\alpha$ and $k_\beta$ as
\begin{align}
\t_{\rm Zak}(k_\alpha,k_\beta)=\sum_{n}^{occ.}\int^{\pi}_{-\pi}[\bm{A}_{n}(\bk)]_\gamma dk_\gamma,
\label{eq_zak}
\end{align}
where $k_\gamma$ is the wave number perpendicular to $k_\alpha$ and $k_\beta$, and the summation is over the occupied states.
At fixed $k_\alpha$ and $k_\beta$, the system is regarded as a one-dimensional insulator.
When the system has $(PT)^2=+$ symmetry or mirror reflection symmetry, the Zak phase quantized as $0$ or $\pi$ $({\rm mod}{~}2\pi)$ \cite{burkov2011topological,fang2015topological,kim2015dirac,yamakage2016line,chan2016ca}.

\section{Topological phase diagram}
\label{sec_numerical}

\begin{figure}
\centering
\includegraphics[width=0.85\hsize]{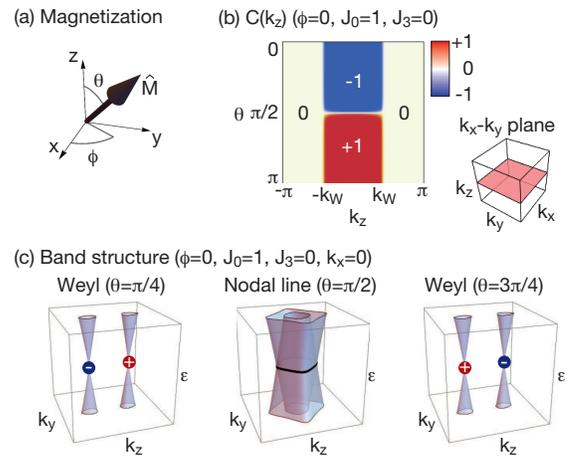}
\caption{(a) Schematic picture of the magnetization $\hat{\bm{M}}$. (b) The Chern number of the $k_x$-$k_y$ plane as a function of $k_z$ and $\t$. The Chern number becomes finite between the Weyl points and its sign is reversed at $\t=\pi/2$. (c) Band structure at $\t=\pi/4$ (Weyl), $\pi/2$ (nodal line), and $3\pi/4$ (Weyl). The Weyl points are depicted with its monopole charge and the nodal line is depicted by a solid line.}
\label{fig_chern}
\end{figure}

In this section, we characterize the topology of the nodal band structure.
In the following calculation, we assume that the energy bands are half filled and we set $a=1$, $t=1$, $m_0=-1/2$, and $m_2=1$ for the sake of simplicity. We numerically calculate Eqs.\ (\ref{eq_chern}) and (\ref{eq_zak}).
Here, we set the magnetization on the $x$-$z$ plane $(\phi=0)$ and the exchange coupling constants as $J_0=1$ and $J_3=0$.
Figure \ref{fig_chern} (b) shows the Chern number of the $k_x$-$k_y$ plane as a function of $k_z$ and $\t$.
In $0\leq\t<\pi/2$, the Chern number is $-1$ for $|k_z|< k_{\rm W}$ where $k_{\rm W}=\arccos(1/4)$ and it becomes zero otherwise. Therefore, there are Weyl points with negative monopole charge at $k_z=-k_{\rm W}$ and with positive monopole charge at $k_z=k_{\rm W}$. In $\pi/2<\t\leq\pi$, on the other hand, the sign of the Chern number is reversed \cite{burkov2018mirror,burkov2018quantum}. The separation of the Weyl points $2k_{\rm W}$ is independent of $\t$.
Figure \ref{fig_chern} (c) shows the nodal band structure. At $\t=\pi/4$ and $3\pi/4$, there is a pair of the Weyl points on the $k_z$ axis. The monopole charge is assigned for each Weyl point and their sign is reversed as we mentioned above. At $\t=\pi/2$, there is a nodal line on the $k_y$-$k_z$ plane. The position of the Weyl points is fixed at $k_z=\pm k_{\rm W}$ and the sign of them are reversed at $\t=\pi/2$ with nodal line structure. The qualitative behavior of the nodal structure is not affected by the $J_3$ term as long as $J_0>J_3$.

\begin{figure*}
\centering
\includegraphics[width=1\hsize]{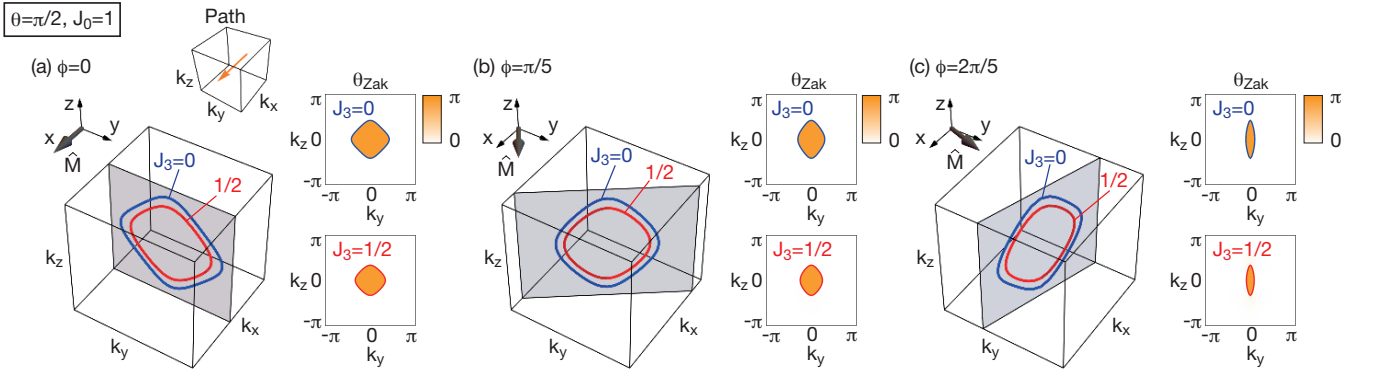}
\caption{
The nodal lines in the Brillouin zone and the Zak phase as a function of $k_y$ and $k_z$ at several azimuth angle $\phi$ of the magnetization. Changing the magnetization direction, the nodal lines rotate and are not gapped as long as the magnetization is in the $x$-$y$ plane. The gray plane represent the plane perpendicular to the magnetization.
The path of integration for calculating the Zak phase is depicted.
When the path of integration passes through the nodal line, the Zak phase is quantized as $\pi$ and it becomes zero otherwise. The qualitative behavior does not change as long as $J_0>J_3$.
}
\label{fig_zak}
\end{figure*}

In the following, we focus on the nodal line at $\t=\pi/2$. 
The magnetization is in the $x$-$y$ plane. 
Figure \ref{fig_zak} shows the nodal line in the Brillouin zone and the Zak phase as a function of $k_y$ and $k_z$.
In Fig.\ \ref{fig_zak} (a), the nodal line resides on the $k_y$-$k_z$ plane at $k_x=0$ for $\phi=0$. The Zak phase is calculated along a path perpendicular to the $k_y$-$k_z$ plane. The Zak phase is quantized as $0$ or $\pi$, due to the mirror-reflection symmetry with respect to the $y$-$z$ plane. 
The Zak phase becomes $\pi$ when the $k_y$ and $k_z$ are set in the area enclosed by the nodal line projected on the $k_y$-$k_z$ plane.
The nodal line shrinks by the orbital-dependent exchange coupling $J_3$ but the qualitative behavior remains unchanged as long as $J_0>J_3$.
In Fig.\ \ref{fig_zak} (b) and (c), the magnetization is tilted from the $x$ axis but the nodal line is not gapped although the mirror-reflection symmetry is broken. 
The Zak phase is quantized and it becomes $\pi$ in the area enclosed by the projected nodal line. This quantization ensures the nodal line structure of the energy bands.
The nodal line almost resides on the plane perpendicular to the magnetization but slightly deviates from the perpendicular plane. Here, we show the nodal lines at $\phi=\pi/5$ and $2\pi/5$ as representative examples. The nodal line structure is retained for arbitrary angle $\phi$.

\begin{figure}
\centering
\includegraphics[width=0.8\hsize]{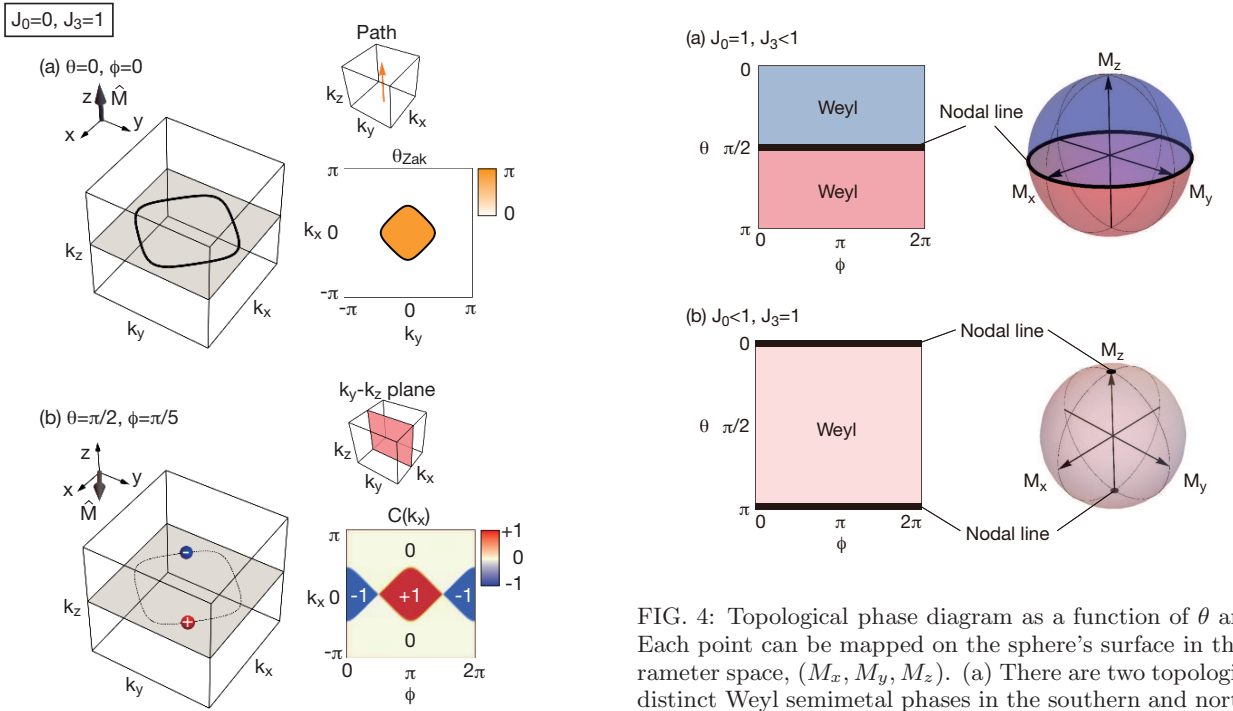}
\caption{The nodal line and the Weyl points in the Brillouin zone. (a) The Zak phase is quantized. (b) There is a pair of the Weyl points and they rotate by changing the magnetization direction. The Chern number on the $k_y$ and $k_z$ plane becomes finite between the Weyl points.}
\label{fig_j3_node}
\end{figure}

Figure \ref{fig_j3_node} shows nodal structure in $J_0=0$ and $J_3=1$.
In this case, nodal structure emerges but the relation between the magnetization direction and the nodal structure is different from the case of $J_0>J_3$.
The nodal line appears on the $k_x$-$k_y$ plane at $k_z=0$ as shown in Fig. \ref{fig_j3_node}(a) when the magnetization is parallel to the $z$ axis.
The Zak phase is calculated along the path perpendicular to the $k_x$-$k_y$ plane and quantized as $0$ or $\pi$.
The nodal line disappears and a pair of Weyl points arises once the magnetization is tilted from the $z$ axis.
Figure \ref{fig_j3_node} (b) shows that the Weyl points in the Brillouin zone at $\phi=\pi/5$ and the Chern number of the $k_y$-$k_z$ plane as a function of $k_x$ and $\phi$.
The dotted line represents the nodal line at $\t=0$ and the Weyl points reside on the dotted line. The $\phi$ dependence of the Chern number shows that the Weyl points move along the dotted line with rotating the magnetization direction. The qualitative behavior is not changed for finite $J_0$ as long as $J_0<J_3$.

\begin{figure}
\centering
\includegraphics[width=0.85\hsize]{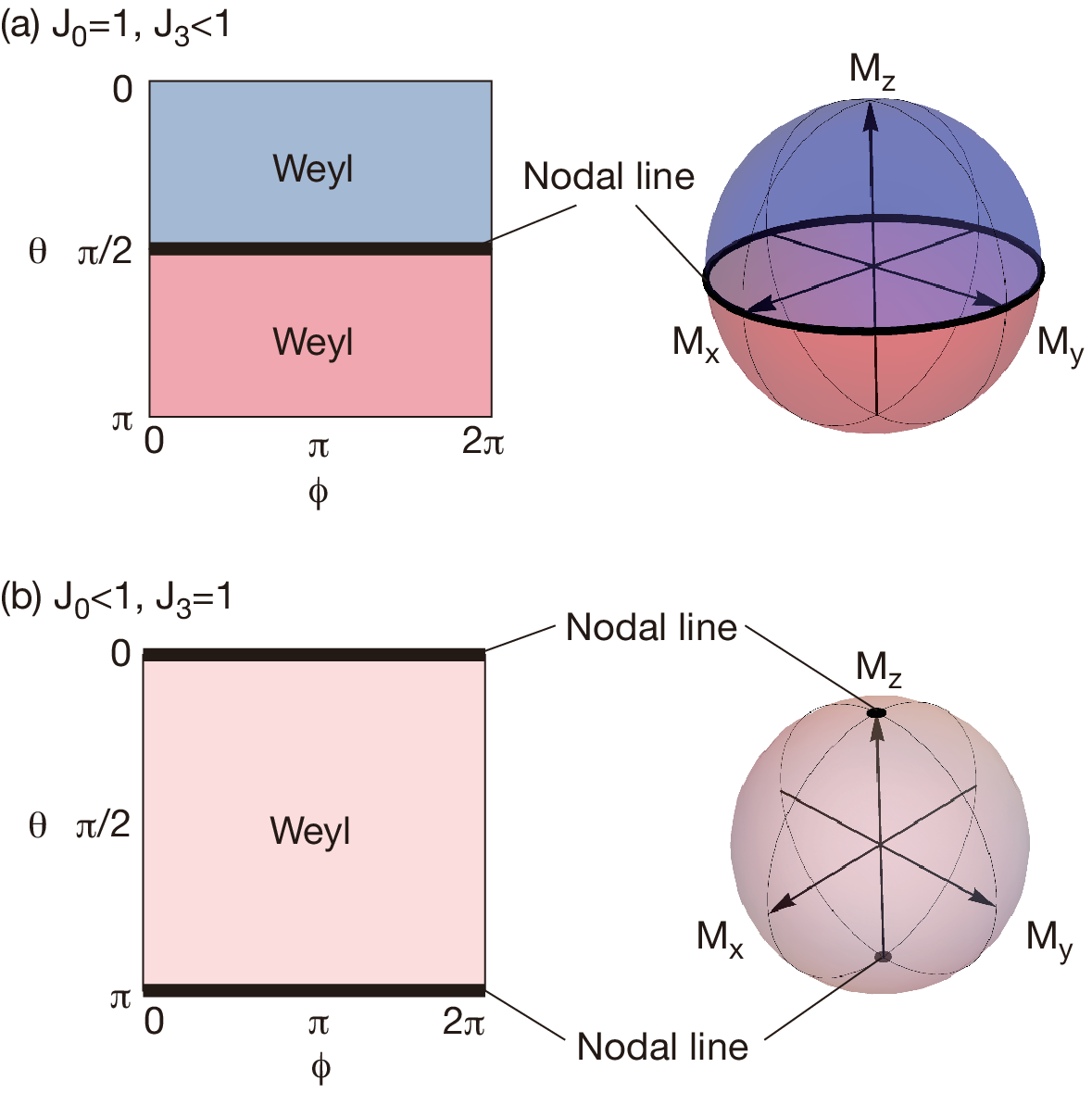}
\caption{Topological phase diagram as a function of $\t$ and $\phi$. Each point can be mapped on the sphere's surface in the parameter space, $(M_x,M_y,M_z)$. (a) There are two topologically distinct Weyl semimetal phases in the southern and northern hemispheres. There is the nodal line semimetal phase on the phase boundary. (b) There is the single Weyl phase. The nodal line phases emerge at the south and north poles.}
\label{fig_phase}
\end{figure}

The phase diagram of the nodal structure is summarized in Fig. \ref{fig_phase}.
The left panels show the phase diagram as a function of $\t$ and $\phi$.
Each point in the left panels is mapped on the unit sphere in $(M_x,M_y,M_z)$ space as shown in the right panels.
The qualitative behavior is different between two cases, $J_0>J_3$ and $J_0<J_3$.
Figure \ref{fig_phase} (a) shows the case of $J_0>J_3$.
There are two topologically distinct Weyl phases in $0\leq\t<\pi/2$ and $\pi/2<\t\leq\pi$.
They are topologically distinguished by the sign of the monopole charge at each Weyl point and one cannot continuously deform the Weyl phase to the other. The nodal line emerges at the equator as a phase boundary between the two Weyl phases.
Figure \ref{fig_phase} (b) shows the case of $J_0<J_3$.
In this case, there is the single Weyl phase. Changing the magnetization direction, the Weyl points move in the momentum space and there is no sign reversal of the Weyl points. However, there are exceptional points at the south pole and the north pole.
When the magnetization direction passes through these points, the signs of the monopole charges are swapped.
These phase diagrams are main results in this work.
We note that there are nodal structures far from the Fermi level. In this work, we focus on the nodal structure close to the Fermi level.


\section{Nodal line for in-plane magnetization}
\label{sec_discussion}

In this section, we discuss the stability of the nodal line in $J_0>J_3$ from the view point of symmetry.
The crystalline symmetry of $H_0$ is compatible to $D_{4h}$.
The generators are given as $C_4=e^{-i\s_z\pi/4}$, $C_2(x)=-i\tau_z\s_x$, and $\s_h=-i\tau_z\s_z$.
The Hamiltonian satisfies the following relations,
\begin{align}
&C_4H_0(k_x,k_y,k_z)C_4^\dagger=H_0(k_y,-k_x,k_z), \\
&C_2(x)H_0(k_x,k_y,k_z)C_2(x)^\dagger=H_0(k_x,-k_y,-k_z), \\
&\s_hH_0(k_x,k_y,k_z)\s_h^\dagger=H_0(k_x,k_y,-k_z).
\end{align}
When the magnetization is in the $x$-$y$ plane, the system has a magnetic reflection symmetry,
\begin{align}
(\s_hT)H(k_x,k_y,k_z)(\s_hT)^{-1}=H(-k_x,-k_y,k_z),
\end{align}
which is equivalent to time reversal symmetry in the two dimensional system at a fixed $k_z$.
In the presence of the magnetic reflection symmetry, the Berry curvature $\Omega_z(\bm k) = \sum_n^{occ.} [\nabla_{\bm k} \times \bm A_{n}(\bm k)] \cdot \hat{\bm z}$ satisfies following relations,
\begin{align}
&\Omega_z(k_x,k_y,k_z)=-\Omega_z(-k_x,-k_y,k_z), 
\label{eq_berry_curvature}
\end{align}
which leads to $C(k_z)=0$.

In the present model, there are nodes on the $k_z$ axis for arbitrary magnetization direction.
On the $k_z$ axis, the Hamiltonian is written as
\begin{align}
&H(0,0,k_z)=m_{k_z}\tau_z+\tau_y\sin k_z+J_0\bm{\s}\cdot\M+J_3\tau_z\bm{\s}\cdot\M,
\end{align}
and the Hamiltonian and the spin operator $\bm{\s}\cdot\M$ commute
\begin{align}
[H(0,0,k_z),\bm{\s}\cdot\M]=0.
\label{spin}
\end{align}
Therefore, the spin is a good quantum number and
the Hamiltonian is written as
\begin{align}
H_\s(k_z)=m_{k_z}\tau_z+\tau_y\sin k_z+\s(J_0+J_3\tau_z),
\end{align}
where $\s=\pm$.
The energy bands are given as
\begin{align}
\e_{\s\pm}(k_z)=\s J_0\pm\sqrt{(m_{k_z}+\s J_3)^2+\sin^2k_z}.
\end{align}
From this expression, one can show the existence of the nodes when the parameters satisfy $J_0>J_3$ and $J_0>|m_0|$.

Using the above relations, we show the stability of the nodal line in the following.
We start with the nodal line in the magnetization along the $x$ axis.
We consider the situation that the magnetization is tilted from the $x$ axis and in the $x$-$y$ plane.
There are nodes on the $k_z$ axis as we mentioned above. If these nodes are Weyl points with monopole charge, the Chern number $C(k_z)$ becomes finite. However, this contradicts Eq.\ (\ref{eq_berry_curvature}). Therefore, the nodal line is stable in the in-plane magnetization.
We note that there are warping terms in the effective model of ${\rm Bi}_2{\rm Te}_3$, which is not included in our model Hamiltonian. The warping terms break the magnetic-reflection symmetry and make the nodal line gapped once the magnetization is tilted from the $x$ axis even in the in-plane magnetization \cite{burkov2018mirror,burkov2018quantum}.

The above property may be lost when anisotropy of the system becomes significant. 
The following exchange coupling is, for instance, possible under the $D_{4h}$-point group symmetry
\begin{align}
 \sigma_x \sin(3\phi) - \sigma_y \cos(3\phi),
\end{align}
which breaks the effective spin conservation law [Eq. (\ref{spin})]. 
This shifts the nodes from the $k_x = k_y = 0$ axis or opens a gap on the nodal line, depending on details of the system. 

\section{conclusion}
\label{sec_conclusion}

We have studied the topological phase diagram in the magnetic topological nodal semimetal.
The topology of band structure can be modulated by changing the direction of the magnetization.
The qualitative behavior is different in $J_0>J_3$ and $J_0<J_3$. In the former case $J_0>J_3$, there are Weyl points on the $k_z$ axis when the magnetization has an out of plane component ($z$-component). When the magnetization is in the $x$-$y$ plane, on the other hand, the nodal line emerges. We found that the nodal line is not gapped as long as the magnetization is in the $x$-$y$ plane. In the present system, there are two topologically distinct Weyl semimetal phases, which are distinguished by the sign of the monopole charge. The nodal line phase can be regarded as a gapless phase between the topologically distinct Weyl semimetal phases. The nodal line is stable as long as the magnetization is in the $x$-$y$ plane. In the latter case $J_0<J_3$, there is the single Weyl semimetal phase. However, the nodal line emerges when the magnetization is parallel to the $z$ axis and the sign reversal of the monopole charge occurs.

\section*{ACKNOWLEDGMENT}
This work was supported by JSPS KAKENHI Grant Numbers JP15H05854, JP17K05485, and JP18H04224, and JST CREST Grant Number JPMJCR18T2.

\bibliography{nodal_line_semimetal}

\end{document}